\documentclass[12pt]{article}
\usepackage{epsfig,amssymb,amsmath,psfrag,subfigure,rotate,color,wasysym}

\allowdisplaybreaks
\newcommand{\insertfig}[2]{\includegraphics[width=#1cm]{#2}}

\def\XXint#1#2#3{{\setbox0=\hbox{$#1{#2#3}{\int}$ }
\vcenter{\hbox{$#2#3$ }}\kern-.6\wd0}}

\def \be  {\begin{equation}}
\def \ee  {\end{equation}}
\def \ba  {\begin{eqnarray}}
\def \ea  {\end{eqnarray}}
\def \baa {\begin{eqnarray*}}
\def \eaa {\end{eqnarray*}}
\def \lab #1 {\label{#1}}

\newcommand\re[1]{(\ref{#1})}
\def\d{\hbox{{d}\kern-.20em\hbox{l}}}

\def \matrix #1 {\left(\begin{array}{cc} #1 \end{array}\right)}

\newcommand \vev [1] {\langle{#1}\rangle}

\newcommand{\ft}[2]{{\textstyle\frac{#1}{#2}}}
\numberwithin{equation}{section}
\begin{document}

\begin{titlepage}

\thispagestyle{empty}

\vspace*{3cm}

\centerline{\large \bf Nonperturbative enhancement of superloop at strong coupling}
\vspace*{1cm}

\centerline{\sc A.V.~Belitsky}

\vspace{10mm}

\centerline{\it Department of Physics, Arizona State University}
\centerline{\it Tempe, AZ 85287-1504, USA}

\vspace{2cm}

\centerline{\bf Abstract}

\vspace{5mm}

We address the near-collinear expansion of NMHV six-particle scattering amplitudes at strong value of the 't Hooft coupling in planar maximally supersymmetric 
Yang-Mills theory. We complement recent studies of this observable within the context of the Pentagon Operator Product Expansion, via the dual superWilson 
loop description, by studying effects of multiple scalar exchanges that accompany (or not) massive flux-tube excitations. Due to the fact that holes have a very small, 
nonperturbatively generated mass $m_{\rm h}$ which is exponentially suppressed in the 't Hooft coupling, their exchanges must be resummed in the ultraviolet limit,  
$\tau \ll m_{\rm h}$. This procedure yields a contribution to the expectation value of the superloop which enters on equal footing with the classical area, --- a phenomenon 
which was earlier observed for MHV amplitudes. In all components, the near-massless scalar exchanges factorize from the ones of massive particles, at leading order in strong 
coupling.

\end{titlepage}

\setcounter{footnote} 0

\newpage

\section{Introduction}

The equivalence between $N$-gluon maximally helicity-violating (MHV) scattering amplitudes in planar maximally supersymmetric gauge theory and the expectation value of the Wilson loop  on 
a null polygonal contour $C_N$ was first established at strong coupling via the analysis of the minimal area in anti-de Sitter space ending on $C_N$ \cite{Alday:2007hr} through the lens of 
gauge/string correspondence \cite{Maldacena:1997re,Witten:1998qj,Gubser:1998bc}. This was further solidified through the Thermodynamic Bethe Ansatz \cite{Alday:2009dv,Alday:2010vh}. 
Simultaneously, extensive perturbative checks verified this duality at weak coupling as well for the MHV case \cite{Drummond:2007cf,Brandhuber:2007yx}. The language suitable for analysis in 
both regimes of weak and strong coupling was recently suggested through the Pentagon Operator Product Expansion \cite{Basso:2013vsa} based on an earlier version \cite{Alday:2010ku}. All-order 
expressions in 't Hooft coupling for the main ingredients of the framework, i.e., the pentagon transitions for all single-particle excitations, including ``flavor'' changing ones, were constructed in a 
series of papers \cite{Basso:2013aha,Belitsky:2014rba,Basso:2014koa,Belitsky:2014sla,Basso:2014nra,Belitsky:2014lta,Belitsky:2015efa,Basso:2014hfa,Basso:2015rta,Belitsky:2015qla} and 
confronted with ``data'' acccumulated in other frameworks to scattering amplitudes at several loop orders 
\cite{Bern:2008ap,DelDuca:2009au,Goncharov:2010jf,Dixon:2013eka,Dixon:2011nj,Dixon:2014iba,Golden:2014xqa,Golden:2014xqf,Golden:2014pua,Drummond:2014ffa,Dixon:2015iva}. 
While the MHV amplitude at strong coupling was addressed in this Operator Product Expansion framework\footnote{Recently scattering matrices that define pentagon transitions were 
independently computed at strong coupling from the perspective of the two-dimensional world-sheet sigma-model in Refs.\ \cite{Bianchi:2015vgw}.} in Refs.\ 
\cite {Basso:2013vsa,Basso:2014koa,Basso:2014jfa,Fioravanti:2015dma} and went beyond the area paradigm in Ref.\ \cite{Basso:2014jfa}, quantitatively not much is known to 
date about the strong coupling behavior of amplitudes at non-MHV level. The latter are dual to a supersymmetric Wilson loop on a null polygonal contour 
\cite{CaronHuot:2010ek,Mason:2010yk,Belitsky:2011zm}. In a recent publication \cite{Belitsky:2015qla}, we had a first glimpse into certain 
components of NMHV hexagon by deriving the inverse-coupling expansion for the pentagons involving gauge fields and fermions. 
However, we have ignored completely contributions due to scalars accompanying any given tree-level exchange that encodes quantum numbers of the transition under study. In the 
present study we will lift this limitation and address the fate of scalar exchanges in NMHV amplitudes. Echoing an earlier work on MHV scattering \cite{Basso:2014jfa}, we will observe 
nonperturbative enhancement of various components due to the nonperturbatively generated hole mass. In fact, we will find that at leading order in the inverse coupling expansion, any given 
component factorizes into the product of terms, one corresponding to the exchange of a massive excitation and the other one due to an infinite number of hole exchanges. Of course, the 
purely scalar components do not admit this factorization. In the current paper, we will focus on the hexagon superloop.

Our subsequent presentation is organized as follows. In the next section, we address the phases of the direct and mirror hole-hole S-matrices and recover their recursive structure
in the non-perturbative scale that allows one to fix the form of the leading contribution to the even and odd parity flux-tube functions. In Sect.\ \ref{NPcorrectionsSection}, we turn to the 
calculation of the first nonperturbative corrections to the latter. Using the hole flux-tube functions, we determine pentagon transitions involving at least one scalar in Sect.\ 
\ref{MixedPentagonsSection}. Then we shift our attention to the application of these results to components of the NMHV hexagon that can accommodate scalars, as the only or an accompanying 
excitation of some transitions. We start with fermionic exchanges and demonstrate the factorizability alluded to above. The same is applicable to the gluonic NMHV exchange as well. We 
perform a resummation of scalar exchanges using numerical studies and a form governed by the interpretation in terms of correlation function of twist operators in O(6) sigma model as was 
done in Ref.\ \cite{Basso:2014jfa} for MHV amplitudes. Along these lines, we find a contribution of the same order as the classical area. Finally we conclude. A couple of appendices 
contain results used in the main text.

\section{Strong-coupling expansion of hole phases}

Let us start our discussion of the strong-coupling regime recalling that the leading contribution of the hole excitation to the expectation value of the Wilson loop arises from its nonperturbative 
regime, i.e., when its rapidity scales as $u \sim O (g^0)$. The solution to the corresponding flux-tube equations, which are quoted for completeness in Appendix \ref{FluxTubeEqsAppendix},
were found in Ref.\ \cite{Basso:2013pxa}. Here we present an indirect way of deducing them. The flux-tube function will not enjoy correct properties,--- it will possess an infinite number of 
poles rather than being an entire function,--- however, the terms which restore its proper analytical structure turn out to be exponentially suppressed in the 't Hooft coupling. The first correction in 
this infinite series will be recovered in the following section.

The indirect method of finding the flux-tube function is based on an iterative structure of scattering phases. It was previously applied in Ref.\ \cite{Basso:2008tx} to the problem of 
nonperturbative corrections to the cusp anomalous dimension. The latter is the vacuum of the flux tube so it should not be surprising that the same formalism is applicable in the current 
circumstances of a hole excitation created on top of the vacuum.

The direct $S_{\rm hh}$ and mirror $S_{\ast \rm hh}$ hole-hole S-matrices, building up the corresponding pentagon transition $P_{\rm h|h}$ \cite{Basso:2013aha}, are determined by 
the dynamical scattering phases $f^{(i)}_{\rm hh}$ \cite{Basso:2013pxa,Belitsky:2014sla} which are integrals of flux-tube hole functions,
\begin{align}
S_{\rm hh} (u_1, u_2) 
&= \exp \left( 2 i \sigma_{\rm hh} (u_1, u_2) - 2i f^{(1)}_{\rm hh} (u_1, u_2) + 2i f^{(2)}_{\rm hh} (u_1, u_2) \right)
\, , \\
S_{\ast \rm hh} (u_1, u_2) 
&= \exp \left( 2 \widehat\sigma_{\rm hh} (u_1, u_2) + 2 f^{(3)}_{\rm hh} (u_1, u_2) - 2 f^{(4)}_{\rm hh} (u_1, u_2) \right)
\, ,
\end{align}
and explicit phases $\sigma_{\rm hh}$ and $\widehat\sigma_{\rm hh}$ that are quoted below in Eqs.\ \re{ExplcitiSigmas}. The strong coupling expansion of $f^{(i)}_{\rm hh}$ will allow us to kill 
two birds with one stone: we will determine the sought after nonperturbative expansion as well as find the leading order flux-tube functions.

Let us start with $f_{\rm hh}^{(1)}$ that can be cast in the form
\begin{align}
\label{PhasefHH1}
f_{\rm hh}^{(1)} (u_1, u_2)
&= \frac{1}{2} \int_0^\infty \frac{dt}{t} \frac{\sin(u_1 t)}{\sinh \frac{t}{2}} \gamma^{\rm h}_{u_2} (2 g t)
\\
&
=
\frac{1}{2}
\int_0^\infty \frac{dt}{t} \sin (u_1 t) \frac{\cosh \frac{t}{2}}{\cosh t} \left[ \Gamma^{\rm h}_{u_2, -} (2 g t) - \Gamma^{\rm h}_{u_2, +} (2 g t) \right]
\nonumber\\
&
+
\frac{1}{2}
\int_0^\infty \frac{dt}{t} \sin (u_1 t) \frac{\sinh \frac{t}{2}}{\cosh t} \left[ \Gamma^{\rm h}_{u_2, -} (2 g t) + \Gamma^{\rm h}_{u_2, +} (2 g t) \right]
\, , \nonumber
\end{align}
making use of a functional transformation \cite{Basso:2008tx,Basso:2008tx1}, see Eq.\ \re{gammaToGamma}, that eliminates explicit dependence  on the coupling constant from 
the flux-tube equations. Performing the inverse Fourier transformation for the product of hyperbolic and trigonometric functions, 
\begin{align}
\label{FT1}
\sin (u t) \frac{\cosh \frac{t}{2}}{\cosh t} 
&
=
- \sqrt{2} g \int_{- \infty}^{\infty} dw \, \sin (2 g t w) \frac{\cosh (g \pi w + u \pi/2)}{\cosh (2 g \pi w + u \pi)} 
\, , \\
\label{FT2}
\sin (u t) \frac{\sinh \frac{t}{2}}{\cosh t} 
&
=
+ \sqrt{2} g \int_{- \infty}^{\infty} dw \, \cos (2 g t w) \frac{\sinh (g \pi w + u \pi/2)}{\cosh (2 g \pi w + u \pi)} 
\, ,
\end{align}
we can rewrite the phase in the form
\begin{align}
f_{\rm hh}^{(1)} (u_1, u_2)
=
&
- \frac{g}{\sqrt{2}}
\int_{- \infty}^{\infty} dw \, \frac{\cosh (g \pi w + u_1 \pi/2)}{\cosh (2 g \pi w + u \pi)} 
\int_0^\infty \frac{dt}{t}
\sin (2 g t w) 
\left[ \Gamma^{\rm h}_{u_2, -} (2 g t) - \Gamma^{\rm h}_{u_2, +} (2 g t) \right]
\nonumber\\
&+
\frac{g}{\sqrt{2}}
\int_{- \infty}^{\infty} dw \, \frac{\sinh (g \pi w + u_1 \pi/2)}{\cosh (2 g \pi w + u \pi)} 
\int_0^\infty \frac{dt}{t}
[ \cos (2 g t w) - 1 ]
\left[ \Gamma^{\rm h}_{u_2, -} (2 g t) - \Gamma^{\rm h}_{u_2, +} (2 g t) \right]
\, .
\end{align}
Here in the second line, we subtracted 1 without any consequences by virtue of Eq.\ \re{FT2} for $t = 0$. Next, splitting the integration range for $w$ into the interval 
$[-1, 1]$ and the rest, we can use the flux-tube equations \re{GammaH1sin} and \re{GammaH2cos} for the former interval, while safely expand integrands at large coupling 
in the latter. After the flux-tube equations had been applied, it remains to evaluate the integrals over the region $[-1,1]$ of $w$, 
\begin{align}
\int_{- 1}^{1} dw \sin (2 g t w) \frac{\cosh (g \pi w + \pi u/2)}{\cosh (2 g \pi w + \pi u)}
&
=
-
\frac{1}{g \sqrt{2}}
\sin (u t)
\frac{\cosh \frac{t}{2}}{\cosh t} 
+ 
\frac{{\rm e}^{- \pi g}}{g} \sinh \frac{u \pi}{2} \, \Re{\rm e} \left[ \frac{{\rm e}^{2 i g t}}{t + i \pi/2} \right]
\nonumber\\
&
+
O ({\rm e}^{- 3 \pi g})
\, , \nonumber\\
\int_{- 1}^{1} dw [ \cos (2 g t w) - 1] \frac{\sinh (g \pi w + \pi u/2)}{\cosh (2 g \pi w + \pi u)}
&
=
\frac{1}{g \sqrt{2}}
\sin (u t)
\frac{\sinh \frac{t}{2}}{\cosh t} 
+ 
\frac{{\rm e}^{- \pi g}}{g} \sinh \frac{u \pi}{2} \, \Re{\rm e} \left[ \frac{i {\rm e}^{2 i g t}}{t + i \pi/2} - \frac{2}{\pi} \right]
\nonumber\\
&
+
O ({\rm e}^{- 3 \pi g})
\, . \nonumber
\end{align}
Adding all of these contributions together, we get
\begin{align}
f_{\rm hh}^{(1)} (u_1, u_2)
&
=
\frac{1}{2}
\int_0^\infty \frac{dt}{t} \frac{\sin (u_1  t)}{\sinh \frac{t}{2}} \left( J_0 (2 g t) - \frac{{\rm e}^{t/2} \cos (u_2 t)}{\cosh t} \right)
\\
&
-
{\rm e}^{- \pi g}
\sinh \frac{u_1 \pi}{2}
\, 
\Re{\rm e}
\Bigg\{
{\rm e}^{i \pi/4}
\int_0^\infty \frac{dt}{t}
\left[
\frac{{\rm e}^{2 i g t}}{t + i \pi/2} + \frac{2i}{\pi}
\right]
\bigg[
i \frac{\cos (u_2 t)}{\sinh \frac{t}{2}}
\nonumber\\
&
\qquad\qquad\qquad\quad
+
\left( 1 + i \coth\frac{t}{2} \right)
\left(
\gamma^{\rm h}_{+, u_2} (2 g t) + i \gamma^{\rm h}_{-, u_2} (2 g t) 
-
J_0 (2 g t)
\right)
\bigg]
\Bigg\}
+
O ({\rm e}^{-3 g \pi})
\, . \nonumber
\end{align}
Comparing this result with Eq.\ \re{PhasefHH1}, we can immediately extract the parity-even flux-tube function
of the hole
\begin{align}
\label{EvengammaHole}
\gamma^{\rm h}_u (2 g t)
=
J_0 (2 g t) - \frac{{\rm e}^{t/2} \cos (u t)}{\cosh t}
+
O ({\rm e}^{-\pi g})
\, ,
\end{align}
up to exponentially-suppressed contributions. Substituting this expression into the $O ({\rm e}^{- \pi g})$ term in the above equation, we find that it vanishes at this order. 
So the first nontrivial correction to the scattering phase will come at order ${\rm e}^{- 2\pi g}$ from the the first nonperturbative term to the flux-tube function. As we pointed 
out earlier and as it is obvious from Eq.\ \re{EvengammaHole}, $\gamma^{\rm h}_u (2 g t)$ possesses an infinite number of fixed poles on the imaginary axis. These are cancelled 
against the ones in nonperturbative terms that we have just mentioned. The first one in this infinite series will be determined in the following section.

To determine $\widetilde\gamma_u^{\rm h}$, we will analyze $f^{(3)}_{\rm hh}$ in the same fashion as above by first changing the basis functions \re{gammaToGammaTilde},
\begin{align}
\label{PhasefHH3}
f_{\rm hh}^{(3)} (u_1, u_2)
&= \frac{1}{2} \int_0^\infty \frac{dt}{t} \frac{\sin(u_1 t)}{\sinh \frac{t}{2}} \widetilde\gamma^{\rm h}_{u_2} (- 2 g t)
\\
&
=
\frac{1}{2}
\int_0^\infty \frac{dt}{t} \sin (u_1 t) \frac{\sinh \frac{t}{2}}{\cosh t} \left[ \widetilde\Gamma^{\rm h}_{u_2, +} (2 g t) - \widetilde\Gamma^{\rm h}_{u_2, -} (2 g t) \right]
\nonumber\\
&
-
\frac{1}{2}
\int_0^\infty \frac{dt}{t} \sin (u_1 t) \frac{\cosh \frac{t}{2}}{\cosh t} \left[ \widetilde\Gamma^{\rm h}_{u_2, +} (2 g t) + \widetilde\Gamma^{\rm h}_{u_2, -} (2 g t) \right]
\, , \nonumber
\end{align}
and then applying the Fourier transforms \re{FT1}, \re{FT2}  with subsequent use of the flux-tube equations \re{GammaHtilde1} and \re{GammaHtilde2}. Then we obtain
\begin{align}
f_{\rm hh}^{(3)} (u_1, u_2)
&
=
\frac{1}{2}
\int_0^\infty \frac{dt}{t} \frac{\sin (u_1  t)}{\sinh \frac{t}{2}} \left( \frac{{\rm e}^{- t/2} \sin (u_2 t)}{\cosh t} \right)
\\
&
-
{\rm e}^{- \pi g}
\sinh \frac{u_1 \pi}{2}
\, 
\Re{\rm e}
\Bigg\{
{\rm e}^{i \pi/4}
\int_0^\infty \frac{dt}{t}
\left[
\frac{{\rm e}^{2 i g t}}{t + i \pi/2} + \frac{2i}{\pi}
\right]
\bigg[
- i \frac{\sin (u_2 t)}{\sinh \frac{t}{2}}
\nonumber\\
&
\qquad\qquad\qquad\quad
+
\left( 1 + i \coth\frac{t}{2} \right)
\left(
\widetilde\gamma^{\rm h}_{+, u_2} (2 g t) - i \widetilde\gamma^{\rm h}_{-, u_2} (2 g t) 
\right)
\bigg]
\Bigg\}
+
O ({\rm e}^{-3 g \pi})
\, . \nonumber
\end{align}
Comparing its right-hand side with Eq.\ \re{PhasefHH3}, we immediately see that this equation defines an iteration for $\widetilde\gamma^{\rm h}_u$ in the perturbative parameter
set by ${\rm e}^{- g \pi}$. Therefore, we find at leading order
\begin{align}
\label{OddgammaHole}
\widetilde\gamma^{\rm h}_{u} (2 g t) = - \frac{\sin (u t) {\rm e}^{t/2}}{\cosh t}
+
O ({\rm e}^{-g \pi})
\, .
\end{align}
Both results for $\gamma^{\rm h}_{u}$ and $\widetilde\gamma^{\rm h}_{u}$ were announced before in Ref.\ \cite{Basso:2013pxa}. Here we obtained them in a rather indirect way as well as 
fixed the form of the first nonperturbative correction to scattering phases. To complete the list of contributing phases, we have to find $f^{(2)}_{\rm hh}$, 
\begin{align}
\label{Initialf2hh}
f_{\rm hh}^{(2)} (u_1, u_2)
=
\int_0^\infty \frac{dt}{t} ({\rm e}^{t/2} \cos (u_1 t) - J_0 (2gt)) \widetilde\gamma^{\rm h}_{u_2} (2 g t)
\, ,
\end{align}
and $f^{(4)}_{\rm hh}$, 
\begin{align}
\label{Initialf4hh}
f_{\rm hh}^{(4)} (u_1, u_2)
=
\int_0^\infty \frac{dt}{t} ({\rm e}^{t/2} \cos (u_1 t) - J_0 (2gt)) \gamma^{\rm h}_{u_2} (- 2 g t)
\, .
\end{align}
The derivation follows the same footsteps. The only difference from the above calculation is the form of the Fourier transform for the integrands, namely, we need
\begin{align}
\label{Fourier1}
\cos (u t) 
\frac{\cosh \frac{t}{2}}{\cosh t}
&
=
\sqrt{2} g \int_{- \infty}^{\infty} dw \cos (2 g w t) \frac{\cosh (g \pi w + u \pi/2)}{\cosh (2 g \pi w + u \pi)}
\, , \\
\label{Fourier2}
\cos (u t) 
\frac{\sinh \frac{t}{2}}{\cosh t}
&
=
\sqrt{2} g \int_{- \infty}^{\infty} dw \sin (2 g w t) \frac{\sinh (g \pi w + u \pi/2)}{\cosh (2 g \pi w + u \pi)}
\, .
\end{align}
Repeating the analysis, we deduce for
\begin{align}
f^{(2)}_{\rm hh} (u_1, u_2)
&
+
\int_0^\infty \frac{dt}{t} (1 - J_0 (2 g t)) \frac{{\rm e}^{t/2} \sin (u_2 t)}{{\rm e}^t - 1}
=
-
\int_0^\infty \frac{dt}{t} \frac{{\rm e}^{t/2} \sin (u_2 t)}{{\rm e}^t - 1} 
\left(
\frac{\cos (u_1 t) {\rm e}^{t/2}}{\cosh t}
-
1
\right)
\nonumber\\
&
+
{\rm e}^{- \pi g}
\cosh \frac{u_1 \pi}{2}
\, 
\Re{\rm e}
\Bigg\{
{\rm e}^{i \pi/4}
\int_0^\infty \frac{dt}{t}
\left[
\frac{{\rm e}^{2 i g t}}{t + i \pi/2} + \frac{2i}{\pi}
\right]
\bigg[
\frac{\sin (u_2 t)}{\sinh \frac{t}{2}}
\\
&\qquad\qquad\qquad\qquad\qquad\quad
+
\left( 1 + i \coth\frac{t}{2} \right)
\left(
\widetilde\gamma^{\rm h}_{+, u_2} (2 g t) - i \widetilde\gamma^{\rm h}_{-, u_2} (2 g t) 
\right)
\bigg]
\Bigg\}
+
O ({\rm e}^{-3 g \pi})
\, , \nonumber
\end{align}
and
\begin{align}
f^{(4)}_{\rm hh} (u_1,u_2) 
&
+
\int_0^\infty \frac{dt}{t} (1 - J_0 (2 g t)) \frac{{\rm e}^{t/2} \cos(u_2 t) - J_0 (2 g t)}{{\rm e}^t - 1} 
\\
&
=
\int_0^\infty \frac{dt}{t ({\rm e}^t - 1)}
\left[
{\rm e}^{t/2} \cos (u_1 t) \left( J_0 (2 g t) - \frac{{\rm e}^{-t/2} \cos (u_2 t)}{\cosh t} \right)
+
\left( {\rm e}^{t/2} \cos (u_2 t) - {\rm e}^t \right) J_0 (2 g t)
\right]
\nonumber\\
&+
{\rm e}^{- \pi g}
\cosh \frac{u_1 \pi}{2}
\, 
\Re{\rm e}
\Bigg\{
{\rm e}^{i \pi/4}
\int_0^\infty \frac{dt}{t}
\left[
\frac{{\rm e}^{2 i g t}}{t + i \pi/2} + \frac{2i}{\pi}
\right]
\bigg[
i \frac{\cos (u_2 t)}{\sinh \frac{t}{2}}
\nonumber\\
&
\qquad\qquad\qquad\quad
+
\left( 1 + i \coth\frac{t}{2} \right)
\left(
\gamma^{\rm h}_{+, u_2} (2 g t) + i \gamma^{\rm h}_{-, u_2} (2 g t) 
-
J_0 (2 g t)
\right)
\bigg]
\Bigg\}
+
O ({\rm e}^{-3 g \pi})
\, , \nonumber
\end{align}
respectively.

Taking the leading order solutions, and adding the explicit phases $\sigma_{\rm hh}$ and $\widehat\sigma_{\rm hh}$,
\begin{align}
\label{ExplcitiSigmas}
\sigma_{\rm hh} (u_1, u_2)
&
=
\int_0^\infty \frac{dt}{t ({\rm e}^t - 1)}
\left[
{\rm e}^{t/2} J_0 (2gt) \sin(u_1 t) - {\rm e}^{t/2} J_0 (2gt) \sin(u_2 t) - {\rm e}^t \sin((u_1 - u_2)t)  
\right]
\, , \\
\widehat\sigma_{\rm hh} (u_1, u_2)
&
=
\int_0^\infty \frac{dt}{t ({\rm e}^t - 1)}
\left[
{\rm e}^{t/2} \left( \cos(u_1 t) + \cos(u_2 t) \right) J_0 (2gt) - \cos((u_1 - u_2)t) - {\rm e}^t J_0^2 (2gt)
\right]
\, ,
\end{align}
the first term in the strong coupling expansion of the hole-hole pentagon reads\footnote{To avoid cluttering the formulas which follow with powers of the 't Hooft coupling,
we normalized the hole-hole pentagon transition, and as a consequence the measure, to coupling independent function at leading order at strong coupling.}
\begin{align}
P_{\rm h|h} (u_1|u_2) 
= 
\frac{\Gamma \left(\ft14 - \ft{i}{4} (u_1 - u_2) \right) \Gamma \left(  \ft{i}{4} (u_1 - u_2) \right)}{4 \Gamma \left(\ft34 - \ft{i}{4} (u_1 - u_2) \right) \Gamma \left( \ft12 + \ft{i}{4} (u_1 - u_2) \right)}
+ \dots
\end{align}
while the measure
\begin{align}
\mu_{\rm h} = \frac{\sqrt{2 \pi^3}}{\Gamma^2 (\ft14)}
+ 
\dots 
\end{align}
is a transcendental constant, with the ellipsis standing for nonperturbative corrections in coupling. The latter can be evaluated with results obtained in the next section. The above 
expressions coincide with the ones derived in Ref.\ \cite{Basso:2014koa}.

\section{Nonperturbative corrections}
\label{NPcorrectionsSection}

Let us now turn to the determination of the exponentially suppressed effects in the flux-tube functions of the hole. As we established in the previous section, the leading order solutions 
yielded functions with incorrect analytical properties. From the point of view of the flux-tube equations with hole inhomogeneities, these generate their particular solutions. We can always 
add homogeneous solutions to the above functions in order to restore analyticity and thus produce an entire function of $t$. As we will find below, these addenda are actually exponentially 
suppressed in the 't Hooft coupling. Below we will provide a recipe for their calculation and construct an explicit first correction to both even and odd parity functions.

\subsection{Even parity}

We start with even parity. Let us add a solution of the homogeneous equation to Eq.\ \re{EvengammaHole}, such that the resulting flux-tube function becomes an entire function in the complex 
$t$-plane,
\begin{align}
\label{PhysicalHgamma}
\gamma^{\rm h}_{u,+} (2 g t) + i \gamma^{\rm h}_{u,-} (2 g t)
=
J_0 (2 g t) + \frac{\sinh \frac{t}{2}}{\sqrt{2} \sinh \left( \frac{t}{2} + i \frac{\pi}{4} \right)}
\left[
\Gamma^{\rm h, \, hom}_{u, +} (2 g t) + i \Gamma^{\rm h, \, hom}_{u, -} (2 g t) - \frac{i \cos (u t)}{\sinh \frac{t}{2}}
\right]
\, .
\end{align}
Presently, we will focus on the cancellation of the leading singularity at $t = - i \pi/2$, however, our consideration can be easily extended to subleading terms as well. This will produce solutions 
to the homogeneous flux-tube equation which induce leading exponential corrections. That is, we impose the following quantization conditions
\begin{align}
\Gamma^{\rm h, \, hom}_{u, +} (4 \pi i x_\ell) + i \Gamma^{\rm h, \, hom}_{u, -} (4 \pi i x_\ell) 
=
- \delta_{\ell, 0} \sqrt{2} \cosh \frac{u \pi}{2}\, ,
\end{align}
where $x_\ell \equiv \ell - \ft14$. A general solution to the homogeneous flux-tube equations was constructed in studies of the flux-tube vacuum \cite{Basso:2008tx1} and reads
\begin{align}
\label{GeneralHomogeneousHGamma}
\Gamma_{u, +}^{\rm h, hom} (\tau) + i \Gamma_{u, -}^{\rm h, hom} (\tau) 
&
=
\sum_{n \geq 1} \frac{c_u^{-} (n, g)}{4 \pi g n - i \tau}
\left[
- i \tau V_0 (- i \tau) U_1^- (4 \pi g n) + 4 \pi g n V_1 (- i \tau) U_0^- (4 \pi g n)
\right]
\nonumber\\
&
+
\sum_{n \geq 1} \frac{c_u^{+} (n, g)}{4 \pi g n + i \tau}
\left[
- i \tau V_0 (- i \tau) U_1^+ (4 \pi g n) + 4 \pi g n V_1 (- i \tau) U_0^+ (4 \pi g n)
\right]
\, ,
\end{align}
where the special functions involved admit the following integral representation
\begin{align*}
V_n (z) 
&
= \frac{\sqrt{2}}{\pi} \int_{-1}^1 dk  \left( \frac{1 + k}{1 - k} \right)^{1/4} \frac{{\rm e}^{k z}}{(1 + k)^n}
\, , \\
U^\pm_n (z) 
&
= \frac{1}{2} \int_{1}^\infty dk  \left( \frac{k + 1}{k - 1} \right)^{\mp1/4} \frac{{\rm e}^{- k (z - 1)}}{(k \mp 1)^n}
\, ,
\end{align*}
and can be related to the confluent hypergemetric function. Substituting these into the quantization conditions and taking the limit $g \to \infty$, making use of their asymptotic 
expansions, which can be found in Refs.\ \cite{Basso:2008tx1,Belitsky:2015qla}, the above quantization conditions can be solved with the result
\begin{align}
c_u^+ (n, g) = - \frac{\Lambda (u, g)}{(8 \pi g n)^{3/4}} \frac{2 \Gamma (n + \ft14)}{\Gamma^2 (\ft14) \Gamma (n)}
\, , \qquad
c_u^- (n, g) = \frac{\Lambda (u, g)}{(8 \pi g n)^{1/4}} \frac{\Gamma (n - \ft14)}{2 \Gamma^2 (\ft34) \Gamma (n)}
\, ,
\end{align}
at leading order in the inverse coupling. Here, we introduced a nonperturbative scale
\begin{align}
\Lambda (u, g) = - \sqrt{2} \cosh \frac{u \pi}{2} \, \frac{{\rm e}^{-\pi g} (2 \pi g)^{5/4}}{\Gamma (\ft54)}
\, .
\end{align}
Substituting these results into Eq.\ \re{GeneralHomogeneousHGamma}, we deduce, after summing the infinite series up,
the leading order contribution to the homogeneous solution of the parity-even flux-tube equation
\begin{align}
\Gamma_{u, +}^{\rm h, hom} (2 g t) + i \Gamma_{u, -}^{\rm h, hom} (2 g t) 
=
\frac{\Lambda (u, g)}{8 \pi g}
\bigg[
&
V_0 (- 2 i g t)
\frac{\Gamma (\ft34) \Gamma (1 - \ft{i t}{2 \pi})}{ \Gamma (\ft34 - \ft{i t}{2 \pi})}
\\
+
&
\left( 2 V_1 (- 2 i g t) - V_0 (- 2 i g t) \right) \frac{\Gamma (\ft54) \Gamma (1 - \ft{i t}{2 \pi})}{ \Gamma (\ft54 - \ft{i t}{2 \pi})}
\bigg]
+
O ({\rm e}^{- 2 \pi g})
\, . \nonumber
\end{align}
This expression can be also recovered from the analysis of Ref.\ \cite{Basso:2008tx1} after appropriate rescaling of nonperturbative vacuum solution.

\subsection{Odd parity}

Let us now turn to odd parity, where the flux-tube function with restored analytical properties is built from \re{OddgammaHole} by adding again a homogeneous solution to it,
\begin{align}
\label{PhysicalHgammaTilde}
\widetilde\gamma^{\rm h}_{u,+} (2 g t) - i \widetilde\gamma^{\rm h}_{u,-} (2 g t)
=
\frac{\sinh \frac{t}{2}}{\sqrt{2} \sinh \left( \frac{t}{2} + i \frac{\pi}{4} \right)}
\left[
\widetilde\Gamma^{\rm h, \, hom}_{u, +} (2 g t) - i \widetilde\Gamma^{\rm h, \, hom}_{u, -} (2 g t) - \frac{\sin (u t)}{\sinh \frac{t}{2}}
\right]
\, .
\end{align}
As above, we will discuss the cancellation of the leading singularity at $t = - i \pi/2$ only. In other words, we impose the following quantization conditions
\begin{align}
\widetilde\Gamma^{\rm h, \, hom}_{u, +} (4 \pi i x_\ell) - i \widetilde\Gamma^{\rm h, \, hom}_{u, -} (4 \pi i x_\ell) 
=
\delta_{\ell, 0} \sqrt{2} \sinh \frac{u \pi}{2}\, ,
\end{align}
where $x_\ell \equiv \ell - \ft14$. Since the homogeneous $\widetilde\Gamma$'s admit the same representation in terms of the infinite series \re{GeneralHomogeneousHGamma}, 
deviating in minor details like certain relative signs, and differ only by the form of the quantization condition, there is no need to redo the analysis anew. We can simply obtain the 
final expression by replacing $\cosh \to - \sinh$ in the even parity solution constructed earlier. The result reads
\begin{align}
\widetilde\Gamma_{u, +}^{\rm h, hom} (2 g t) - i \widetilde\Gamma_{u, -}^{\rm h, hom} (2 g t) 
=
\frac{\widetilde\Lambda (u, g)}{8 \pi g}
\bigg[
&
V_0 (- 2 i g t)
\frac{\Gamma (\ft34) \Gamma (1 - \ft{i t}{2 \pi})}{ \Gamma (\ft34 - \ft{i t}{2 \pi})}
\\
+
&
\left( 2 V_1 (- 2 i g t) - V_0 (- 2 i g t) \right) \frac{\Gamma (\ft54) \Gamma (1 - \ft{i t}{2 \pi})}{ \Gamma (\ft54 - \ft{i t}{2 \pi})}
\bigg]
+
O ({\rm e}^{- 2 \pi g})
\, , \nonumber
\end{align}
with
\begin{align}
\widetilde\Lambda (u, g) = \sqrt{2} \sinh \frac{u \pi}{2} \, \frac{{\rm e}^{-\pi g} (2 \pi g)^{5/4}}{\Gamma (\ft54)}
\, .
\end{align}

We can verify the correctness of these expressions by substituting them into Eq.~\re{PhysicalHgamma} and calculating the energy and momentum of the hole excitation
\re{HoleEandPfromGammaH}. We find
\begin{align}
E_{\rm h} (u) 
=
- \frac{\Lambda (u, g)}{2 \pi g} = m_{\rm h} \cosh \frac{u \pi}{2}
\, , \qquad
p_{\rm h} (u) 
= 
\frac{\widetilde\Lambda (u, g)}{2 \pi g} = m_{\rm h} \sinh \frac{u \pi}{2}
\, , 
\end{align}
which is in agreement with the well-known leading order result \cite{Basso:2010in} when expressed in terms of the nonperturbatively generated mass of the O(6) sigma 
model \cite{Alday:2007mf,Basso:2008tx}
\begin{align}
m_{\rm h} = {\rm e}^{- \pi g} \frac{(8 \pi g)^{1/4}}{\Gamma (\ft54)} + O ({\rm e}^{-2 \pi g})
\, .
\end{align}
The above consideration can be extended to higher orders without facing conceptual difficulties.

\section{Mixed pentagons}
\label{MixedPentagonsSection}

With the found explicit expressions for the hole flux-tube functions, we can determine the mixed hole-fermion and hole-gluon scattering phases at strong coupling.
The only integral that one needs for the leading order solution is the following one
\begin{align*}
\int_0^\infty \frac{dt}{t} \frac{{\rm e}^{i \alpha t} - 1}{{\rm e}^t + 1} = \ln \frac{\Gamma \left( \ft12 - i \ft{\alpha}{2} \right)}{\sqrt{\pi} \Gamma (1 - i \ft{\alpha}{2})}
\, .
\end{align*}
Notice that while hole's rapidity will stay in the nonpertubative domain $u_{\rm h} \sim O (g^0)$, the ones for the fermion and gauge field should belong to the
perturbative strong coupling scaling regime, where their energy and momentum are of order $g^0$,
\begin{align}
E_\star  = m_\star \cosh\theta
\, , \qquad
p_\star = m_\star \sinh\theta
\end{align}
with $m_{\rm f} = 1$ and $m_{\rm g} = \sqrt{2}$ \cite{Alday:2007mf}, to bestow amplitudes with leading contributions. Thus, the fermion belongs to the small fermion sheet with 
rapidity $u_{\rm f} = 2 g \widehat{u}_{\rm f}$ where $|\widehat{u}_{\rm f}| = |\coth( 2 \theta) | > 1$, while the gluon one scales as $u_{\rm g} = 2 g \widehat{u}_{\rm g}$ with 
$|\widehat{u}_{\rm g}| = |\tanh (2 \theta)| < 1$.

The hole-small fermion phases are
\begin{align}
&
f_{\rm hf}^{(1)} (u, v) = - \frac{1}{16 g^2} \frac{u}{\widehat{v}^2} + \dots
\, , \quad
f_{\rm hf}^{(2)} (u, v) = \frac{1}{8 g} \frac{1}{\widehat{v}} + \dots
\, , \\
&
f_{\rm hf}^{(3)} (u, v) = - \frac{1}{4 g} \frac{u}{\widehat{v}} + \dots
\, , \qquad \ 
f_{\rm hf}^{(4)} (u, v) = - \frac{1}{2} \ln \big( 2 \widehat{v} \widehat{x}_{\rm f}[v] \big) + \frac{1}{4 g^2} \left( \frac{3}{16} + \frac{u^2}{4} \right) \frac{1}{\widehat{v}^2} + \dots
\, , \nonumber
\end{align}
where the rescaled small-fermion Zhukowski variable in the hyperbolic parametrization reads $\widehat{x}_{\rm f} = \tanh (\theta)$. Such that the the hole-small fermion pentagon in the 
regime in question takes the form at leading order
\begin{align}
\label{PhfLO}
P_{\rm h|f} (u|v) = \frac{1}{\sqrt{2 g \widehat{v}}} \exp\left( \frac{u^-}{2 g \widehat{v}}+ \dots \right)
\, .
\end{align}
We can immediately test this form by employing constraints stemming from the $\bar{Q}$-equation \cite{CaronHuot:2011kk,Bullimore:2011kg}, namely, as demonstrated in
Ref.\ \cite{Belitsky:2015kda}, it enters the following integral equation
\begin{align}
\int d v \, \mu_{\rm f} (v) {\rm e}^{- \tau (E_{\rm f} (v) - 1)} x_{\rm f}^{3/2} [v] \delta \left( p_{\rm f} (v) \right) P_{\rm \bar{f}|f} (- u + \ft{3i}{2} | v) P_{\rm h|f} (- u | v) = \frac{2 g^3}{\Gamma (g)}
\, ,
\end{align}
where $\Gamma (g)$ is the cusp anomalous dimension. Rescaling the small-fermion rapidity and re-expressing it in terms of the Zhukowski variable $\widehat{u}_{\rm f} = \widehat{x}_{\rm f}
+ 1/\widehat{x}_{\rm f}$, one can immediately confirm the leading order expression for the hole-small fermion pentagon \re{PhfLO} making use of the known expressions for the small 
fermion measure and fermion-antifermion pentagon \cite{Basso:2014koa}
\begin{align}
\mu_{\rm f} (u) = - \left( 1 - \widehat{x}_{\rm f}^2[u] \right)^{-1/2} + \dots
\, , \qquad
P_{\rm \bar{f}|f} (u| v) = \left( 1 - \widehat{x}_{\rm f} [u] \widehat{x}_{\rm f}[v] \right)^{-1/2} + \dots 
\, .
\end{align}

For the gluon-hole case, it is more instructive to discuss the entire direct and mirror S-matrices rather than individual phases. Then one
finds by substituting the leading order hole solutions \re{EvengammaHole} and \re{OddgammaHole} to the dynamical phases that they cancel exactly the $\sigma$-phases on the level
of integrands and thus both S-matrices are trivial 
\begin{align}
S_{\rm hg} = 1
\, , \qquad
S_{\rm \ast hg} = 1
\, ,
\end{align} 
up to nonperturbative effects in coupling. Consequently, the hole-gluon pentagon is
\begin{align}
\label{HoleGluonPentagon}
P_{\rm h|g} (u| v) = 1 + O ({\rm e}^{- \pi g})
\, .
\end{align}
With these results at our disposal, we are now ready to move onto explicit analyses of different components of the NMHV hexagon at strong coupling.

\section{Hexagon superloop}

Let us decompose the hexagon superloop at the NMHV level in terms of Grassmann components that receive leading contribution from single
particle exchanges,
\begin{align}
\label{NMHVhexagon}
\mathcal{W}_6 (\tau, \sigma, \phi) 
&
= 
\chi_1^4 {\rm e}^{i \phi} \mathcal{W}_{(4,0)} (\tau, \sigma)
\\
&
+ 
\chi_1^3 \chi_4 
\left( 
{\rm e}^{i \phi/2} \mathcal{W}^{\rm odd}_{(3,1)} (\tau, \sigma) 
+  
{\rm e}^{- i \phi/2} \mathcal{W}^{\rm even}_{(3,1)} (\tau, \sigma)
\right) 
+ 
\chi_1^2 \chi_4^2 \mathcal{W}_{(2,2)} (\tau, \sigma)
+
\dots \, . \nonumber
\end{align}
Here we adopted a conventional twistor parametrization via the three variables $\tau$, $\sigma$ and $\phi$ which are equivalent to the three conformal cross-ratios $u$, $v$ and $w$ 
of the six-point remainder function. We will start below with the second contribution $\mathcal{W}_{(3,1)}$ in the Grassmann series, the one that is induced by the (anti)fermion production on the 
bottom and absorption at the top along with an infinite number of scalars. We divided their effect in the above sum in two classes, an antifermion along with an even number of scalars 
and a fermion with an odd number of scalars. As a consequence, $\mathcal{W}^{\rm odd}_{(3,1)}$ starts with two-particle exchanges compared to $\mathcal{W}^{\rm even}_{(3,1)}$. Both of 
them transform in the ${\bf \bar 4}$ of SU(4), however, possess different helicities as exhibited by accompanying phases in the above equation.

\subsection{Antifermion-scalars}

The leading contribution to $\mathcal{W}^{\rm even}_{(3,1)}$ comes from the single-particle exchange with the quantum numbers of the antifermion
\begin{align}
\label{Wf-bar}
\mathcal{W}_{\rm \bar{f}} =  \int d \mu_{\rm f} (v) x_{\rm f} [v]
\, ,
\end{align}
with the NMHV helicity form factor determined by the small-fermion Zhukowski variable $x_{\rm f} [v] = \ft12 (v - \sqrt{v^2 - (2 g)^2})$. Here and below the single particle measure includes
the propagating phases
\begin{align}
d \mu_\star (v) = \frac{d v}{2 \pi} \mu_\star (v) {\rm e}^{- \tau E_\star (v) + i \sigma p_\star (v)}
\end{align}
that are determined by all-order energies $E_\star$ and momenta $p_\star$ \cite{Basso:2010in}. 
Next in the infinite series comes the antifermion accompanied by two scalars
\begin{align}
\mathcal{W}^{\rm\bf \bar{4}}_{\rm hh\bar{f}} 
=
\frac{1}{2!}
\int d \mu_{\rm h} (u_1) d \mu_{\rm h} (u_2)
\int d \mu_{\rm f} (v) x_{\rm f} [v] \frac{1}{|P_{\rm h|f} (u_1| v) P_{\rm h|f} (u_2| v)|^2}
\frac{\Pi^{\rm\bf \bar{4}}_{\rm hhf} (u_1, u_2, v)}{|P_{\rm h|h} (u_1| u_2)|^2}
\, ,
\end{align}
where the matrix part reads
\begin{align}
\Pi^{\rm\bf \bar{4}}_{\rm hhf} (u_1, u_2, v)
=
\frac{3}{2}
\frac{45 + 6 u_1^2 - 8 u_1 u_2 + 6 u_2^2 - 4 (u_1 + u_2) v + 4 v^2
}{
[1 + (u_1 - u_2)^2] [4 + (u_1 - u_2)^2] [\ft{9}{4} + (u_1 - v)^2] [\ft{9}{4} + (u_2 - v)^2]
}
\, .
\end{align}
In the scaling limit $v = 2 g \widehat{v}$ with $g \to \infty$ and $\widehat{v} =  {\rm fixed}$, we get
\begin{align}
\Pi^{\rm\bf \bar{4}}_{\rm hhf} (u_1, u_2, v) = \frac{1}{v^2} \Pi^{\rm\bf 1}_{\rm hh} (u_1, u_2) + O (1/v)
\, ,
\end{align}
with $\Pi^{\rm\bf 1}_{\rm hh}$ being the singlet two-scalar matrix part (which defines one of the twist-two contributions in the MHV amplitude \cite{Basso:2014koa})
\begin{align}
\Pi^{\rm\bf 1}_{\rm hh} (u_1, u_2)
=
\frac{6
}{
[1 + (u_1 - u_2)^2] [4 + (u_1 - u_2)^2]
}
\, .
\end{align}
This immediately yields a factorized form of the three-particle contribution in terms of the fermion, on the one hand, and the two-scalar pair in the singlet representation, on the 
other, i.e., 
\begin{align}
\mathcal{W}^{\rm\bf \bar{4}}_{\rm hh\bar{f}} = \mathcal{W}_{\rm \bar{f}}  \mathcal{W}^{\rm\bf 1}_{\rm hh}
\end{align}
where
\begin{align}
\mathcal{W}^{\rm\bf 1}_{\rm hh}
=
\frac{1}{2!}
\int d \mu_{\rm h} (u_1) d \mu_{\rm h} (u_2)
\frac{\Pi^{\rm\bf 1}_{\rm hh} (u_1, u_2)}{|P_{\rm h|h} (u_1| u_2)|^2}
\, .
\end{align}
A simple counting of the powers of the 't Hooft coupling demonstrates that $\mathcal{W}^{\rm\bf 1}_{\rm hh}$ is of order $g^0$ and contributes on equal footing with 
Eq.\ \re{Wf-bar}. The same phenomenon persists for all multi-scalar exchanges such that all scalar pairs have to be accounted for,
\begin{align}
\label{WevenAllOrder}
\mathcal{W}^{\rm even}_{(3,1)}
=
\mathcal{W}_{\rm \bar{f}}
\sum_{n = 0}^\infty \, \mathcal{W}^{\rm\bf 1}_{({\rm hh})^n}
\, ,
\end{align}
with $\mathcal{W}^{\rm\bf 1}_{({\rm hh})^0} = 1$ and $\mathcal{W}^{\rm\bf 1}_{({\rm hh})^n}$ having the form analogous to the two-scalar contribution 
\begin{align}
\label{EvenMultiScalars}
\mathcal{W}^{\rm\bf 1}_{({\rm hh})^n}
=
\frac{1}{(2 n)!}
\int d \mu_{\rm h} (u_1) \dots d \mu_{\rm h} (u_{2n})
\frac{\Pi^{\rm\bf 1}_{\rm h \dots h} (u_1, \dots , u_{2n})}{ \prod_{i<j}^{2n} |P_{\rm h|h} (u_i| u_j)|^2}
\end{align}
with the matrix part $\Pi^{\rm\bf 1}_{\rm h \dots h} (u_1, \dots , u_{2n})$ that can be read off from the integral representation given in Refs.\ \cite{Basso:2014jfa,Basso:2015uxa}.

\subsection{Singlet multi-scalar exchanges}

\begin{figure}[t]
\begin{center}
\mbox{
\begin{picture}(0,220)(240,0)
\put(0,10){\insertfig{7}{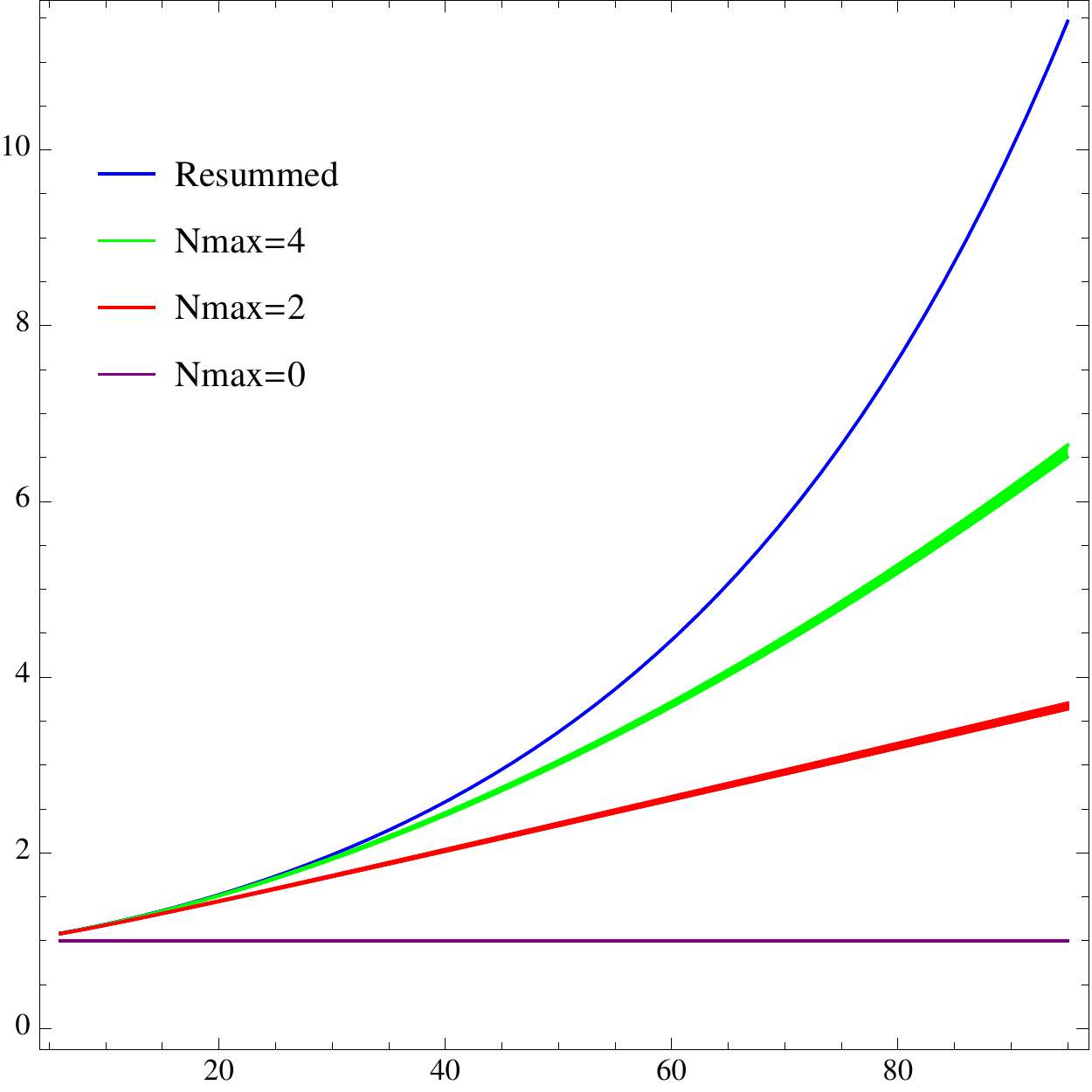}}
\put(278,10){\insertfig{7}{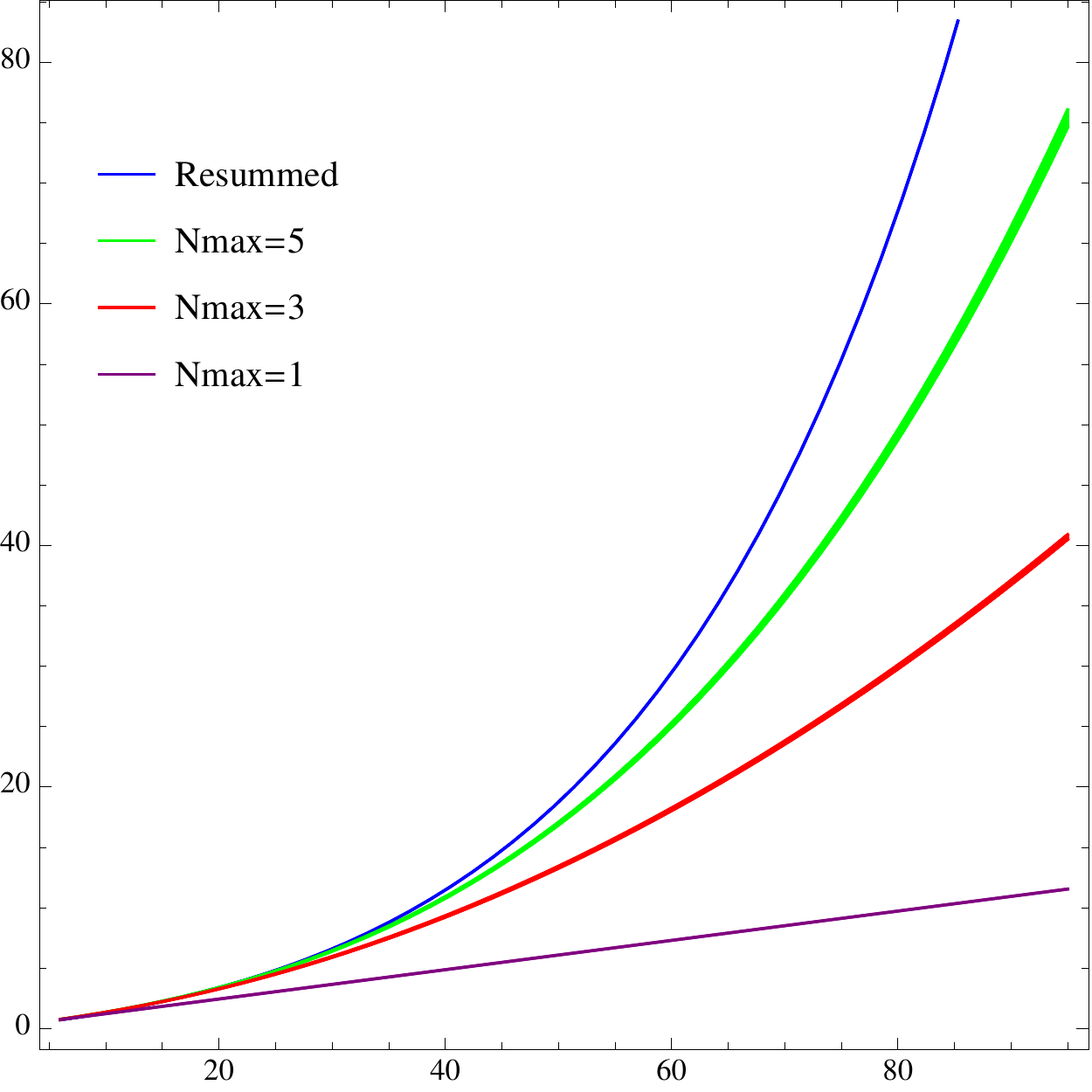}}
\put(16,190){\tiny $\mathcal{W}^{\bf 1}_\infty \, \mbox{vs} \, \ln (m_{\rm h} \xi)^{-1}$}
\put(295,190){\tiny $\mathcal{W}^{\bf 6}_\infty/g \, \, \mbox{vs} \, \ln (m_{\rm h} \xi)^{-1}$}
\put(95,-7){(a)}
\put(375,-7){(b)}
\end{picture}
}
\end{center}
\caption{ \label{Hseries} Plots of the truncated at $N_{\rm max}$ series of hole contributions to the singlet \re{SingletSsm} and sextet \re{ResummedSextet} Wilson loops,
shown in (a) and (b)panels, respectively. Panel (a) displays the effect of adding successive terms for $N_{\rm max} = 0, 2, 4$, while panel (b) shows it for $N_{\rm max} = 1, 3, 5$. 
The resummed curves correspond to the right-hand sides of Eqs.\ \re{SingletSsm} and \re{ResummedSextet}, respectively. All curves were obtained by averaging Monte Carlo 
computations of integrals at points $m_{\rm h} \xi = 10^{-3},10^{-6}, 10^{-10}, 10^{-14}, 10^{- 18}, 10^{-25}, 10^{-40}$ and fitting the outcomes. The thickness of the curves shows 
one standard deviation of Monte Carlo data.}
\end{figure}

In spite of the fact that the singlet multi-scalar resummation was analyzed in Ref.\ \cite{Basso:2014jfa}, in preparation for the sextet case that is addressed next, we will 
repeat numerical computations here and confront them against twist-operator correlation functions. We start with the two-particle contribution. In the ultraviolet regime, it has 
the following asymptotic form
\begin{align}
\mathcal{W}^{\rm\bf 1}_{\rm hh}  |_{m_{\rm h} \xi \ll 1}
=
\mu_{\rm h}^2
\left[
\alpha^{\rm\bf 1}_{\rm hh} \ln \frac{1}{m_{\rm h} \xi} + \beta^{\rm\bf 1}_{\rm hh} \ln \ln \frac{1}{m_{\rm h} \xi} + \gamma^{\rm\bf 1}_{\rm hh} 
\right]
+ 
O \left( (m_{\rm h} \xi)^0 \right)
\, ,
\end{align}
where the relativistic invariance of the contributions is exhibited through the dependence on a single variable $\xi = \sqrt{\tau^2 + \sigma^2}$. The coefficients accompanying
functional dependence on $\xi$ can be partially determined analytically
\begin{align}
\alpha^{\rm\bf 1}_{\rm hh}
&
=
\frac{\Gamma^4 (\ft14)}{84 \pi^5} 
\left[
56 + 5 \, {_4{F}_3} \left.\left( {1, 1, \ft54, \ft94 \atop 2, \ft74, \ft{11}4} \right| 1 \right)
\right]
-
\frac{1}{5 \pi^3}
\left[
40
+ 3 \pi \,
{_3{F}_2} \left.\left( {\ft12, \ft54, \ft32 \atop 2, \ft94} \right| 1 \right)
\right]
\nonumber\\
&
\simeq
0.087
\, , \\
\beta^{\rm\bf 1}_{\rm hh}
&
\simeq
- 0.137 \pm 0.001
\, ,\\
\gamma^{\rm\bf 1}_{\rm hh}
&
\simeq
- 0.044 \pm 0.014
\, .
\end{align}
The effects of subleading terms in the expansion were analyzed numerically. In Fig.\ \ref{Hseries} (a), we demonstrate the result of successive additions of more and more
scalar exchanges in Eq.\ \re{WevenAllOrder}. In these estimates, we computed the multifold integrals \re{EvenMultiScalars} making use of an adaptive Monte Carlo method 
and then averaged over multiple samples. The standard deviation from the mean is shown in above formulas, while it is stripped off the graphs as not to obscure effects 
from multi-hole exchanges. We confirmed quick convergence of the Operator Product Expansion in the range of $m_{\rm h} \xi < 10^{-18}$ in this channel with the resulting 
functional fit inspired by the two-point correlation function of twist operators $\phi_{\pentagon}$, whose matrix elements correspond to the pentagon transitions as was pointed
out in \cite{Basso:2014jfa},
\begin{align}
\label{SingletSsm}
\mathcal{W}^{\rm\bf 1}_\infty
\equiv
\sum_{n = 0}^\infty \, \mathcal{W}^{\rm\bf 1}_{({\rm hh})^n}  |_{m_{\rm h} \xi \ll 1}
=
C^{\rm\bf 1}_{\rm h} \, (m_{\rm h} \xi)^{-1/36} \ln^{-1/24} \frac{1}{m_{\rm h} \xi}
\, ,
\end{align}
and $C^{\rm\bf 1}_{\rm h} \simeq 0.99$. This is indeed the result of Ref.\ \cite{Basso:2014jfa}. We will take it below as a basis for our analysis of the sextet component in the 
hexagonal Wilson loop.

\subsection{Fermion-scalars}

The even-scalar exchanges do not have ${\bf 6}$ in their product, e.g., ${\bf 6} \times {\bf 6} = {\bf 1} + {\bf 15} + {\bf 20}$, so they do not contribute to the component 
$\mathcal{W}^{\rm odd}_{(3,1)}$ in question. However, any odd number of holes does contribute. The first term in the fermion-scalar series is the one with a fermion
and a hole in the ${\bf \bar 4}$ of SU(4), emerging from the product ${\bf 4} \times {\bf 6} = {\bf \bar 4} + {\bf 20}$,
\begin{align}
\mathcal{W}^{\bf \bar{4}}_{\rm hf} = \int d \mu_{\rm h} (u) \int d \mu_{\rm f} (v) 
\frac{\Pi^{\bf \bar{4}}_{\rm hf} (u, v) }{|P_{\rm h|f} (u | v)|^2}
\, ,
\end{align}
with
\begin{align}
\Pi^{\bf \bar{4}}_{\rm hf} (u, v) = \frac{3}{ [(u - v)^2 + \ft{9}{4}] }
\, .
\end{align}
The leading contribution at strong coupling emerges from the scaling limit of the small fermion rapidity $v = 2 g \widehat{v}$ with $\widehat{v} \sim O (g^0)$ and 
nonperturbative one for the hole, i.e., $u \sim O (g^0)$. Then the matrix part immediately simplifies and we get
\begin{align}
\mathcal{W}^{\bf \bar{4}}_{\rm f h}  = 3 W_{\rm f} W^{\bf 6}_{\rm h}
\end{align}
with
\begin{align}
W_{\rm f} =  \int d \mu_{\rm f} (\widehat{v}) \frac{\widehat{x}_{\rm f}[v]}{(1 + \widehat{x}^2_{\rm f}[v])} 
\, , \qquad
W^{\bf 6}_{\rm h} = g \int d \mu_{\rm h} (u)
\, .
\end{align}
We pulled out the factor of 3 stemming from SU(4) tensor contraction as an overall coefficient. The next term arises from the four-particle ${\rm hhhf}$-state
\begin{align}
\mathcal{W}^{\bf \bar{4}}_{\rm hhh\bar{f}} 
=
\frac{1}{3!}
\int d \mu_{\rm h} (u_1) d \mu_{\rm h} (u_2) d \mu_{\rm h} (u_3)
&
\int d \mu_{\rm f} (v) \frac{1}{|P_{\rm h|f} (u_1| v) P_{\rm h|f} (u_2| v) P_{\rm h|f} (u_3| v)|^2}
\nonumber\\
&\times
\frac{\Pi^{\bf \bar{4}}_{\rm hhhf} (u_1, u_2, u_3, v)}{|P_{\rm h|h} (u_1| u_2) P_{\rm h|h} (u_1| u_3) P_{\rm h|h} (u_2| u_3)|^2}
\, ,
\end{align}
where $\Pi^{\bf \bar{4}}_{\rm hhhf}$ is too cumbersome to be displayed here. However, in the scaling limit, it reduces to 
\begin{align}
\label{FactorizationPiSix}
\Pi^{\bf \bar{4}}_{\rm hhhf} (u_1, u_2, u_3, v)
=
\frac{3}{v^4} \Pi^{\rm\bf 6}_{\rm hhh} (u_1, u_2, u_3) + O (v^{-5})
\, , 
\end{align}
where the matrix part of the three-hole state in the ${\bf 6}$ of SU(4) reads
\begin{align}
&
\Pi^{\rm\bf 6}_{\rm hhh} (u_1, u_2, u_3)
\\
&
=
6
\frac{
[7 + u_1^2 + u_2^2 + u_3^2 - (u_1 u_2 + u_1 u_3 + u_2 u_3)] 
[12 + u_1^2 + u_2^2 + u_3^2 - (u_1 u_2 + u_1 u_3 + u_2 u_3)]
}
{
[1 + (u_1 - u_2)^2] [4 + (u_1 - u_2)^2] [1 + (u_1 - u_3)^2] [4 + (u_1 - u_3)^2] [1 + (u_2 - u_3)^2] [4 + (u_2 - u_3)^2]
}
\, . \nonumber
\end{align}
Therefore, the expression factorizes again yielding
\begin{align}
\mathcal{W}^{\bf \bar{4}}_{\rm hhh\bar{f}} 
=
3 W_{\rm f} W^{\bf 6}_{\rm hhh}
\end{align}
with
\begin{align}
W^{\bf 6}_{\rm hhh}
=
\frac{g}{3!}
\int d \mu_{\rm h} (u_1) d \mu_{\rm h} (u_2) d \mu_{\rm h} (u_3)
\frac{\Pi^{\rm\bf 6}_{\rm hhh} (u_1, u_2, u_3)}{|P_{\rm h|h} (u_1| u_2) P_{\rm h|h} (u_1| u_3) P_{\rm h|h} (u_2| u_3)|^2}
\, .
\end{align}
Using the integral representation of the matrix part of the pentagon transitions \cite{Basso:2015uxa}, one can convince oneself that the above property \re{FactorizationPiSix} 
persists for any odd number of hole excitations, such that any number of scalars accompanying the fermion needs to be resumed
\begin{align}
\mathcal{W}^{\rm odd}_{(3,1)}
=
3 \mathcal{W}_{\rm f}
\sum_{n = 0}^\infty \, \mathcal{W}^{\rm\bf 6}_{{\rm h}({\rm hh})^n}
\, ,
\end{align}
where similarly to Eq.\ \re{EvenMultiScalars}
\begin{align}
\label{SextetSsm}
\mathcal{W}^{\rm\bf 6}_{{\rm h} ({\rm hh})^n}
=
\frac{g}{(2 n + 1)!}
\int d \mu_{\rm h} (u_1) \dots d \mu_{\rm h} (u_{2n + 1})
\frac{\Pi^{\rm\bf 6}_{\rm h \dots h} (u_1, \dots , u_{2n + 1})}{ \prod_{i<j}^{2n + 1} |P_{\rm h|h} (u_i| u_j)|^2}
\end{align}
with the matrix part of sextet hole exchanges that can be read off from the integral representation given in Refs.\ \cite{Basso:2014jfa,Basso:2015uxa}.

\subsection{Sextet multi-scalar exchanges}

The one-particle contribution with sextet quantum numbers arises from the single hole exchange
\begin{align}
\mathcal{W}^{\rm\bf 6}_{\rm h} 
&
=  g \int d \mu_{\rm h} (u)
= g \frac{2 \mu_{\rm h}}{\pi^2} K_0 (m_{\rm h} \xi)
\, , \nonumber
\end{align}
where after the second equality sign we displayed its leading behavior from nonperturbative domain of rapidities. In the infrared regime $m_{\rm h} \xi \gg 1$, it displays the expected 
exponentially suppressed behavior
\begin{align}
\mathcal{W}^{\rm\bf 6}_{\rm h} |_{m_{\rm h} \xi \gg 1} 
=
g \frac{\sqrt{2}\mu_{\rm h}}{\pi^{3/2}}
\frac{{\rm e}^{- m_{\rm h} \xi}}{\sqrt{m_{\rm h} \xi}} \, \left( 1 + O \big( 1/(m_{\rm h} \xi) \big) \right)
\, ,
\end{align}
while in the ultraviolet region $m_{\rm h} \xi \ll 1$, it shows logarithmic enhancement,
\begin{align}
\mathcal{W}^{\rm\bf 6}_{\rm h} |_{m_{\rm h} \xi \ll 1} = g \frac{2 \mu_{\rm h}}{\pi^2} \ln \frac{1}{m_{\rm h} \xi} + O \big( (m_{\rm h} \xi)^0 \big)
\, .
\end{align}

The enhancement of the ultraviolet regime persists and amplifies in multi-hole exchanges. For instance, in the three-particle term that reads
\begin{align}
W^{\rm\bf 6}_{\rm hhh}
=
\frac{g}{3!}
\int d \mu_{\rm h} (u_1) d \mu_{\rm h} (u_2) d \mu_{\rm h} (u_3)
\frac{\Pi^{\rm\bf 6}_{\rm hhh} (u_1, u_2, u_3)}{|P_{\rm h|h} (u_1| u_2) P_{\rm h|h} (u_1| u_3) P_{\rm h|h} (u_2| u_3)|^2}
\, ,
\end{align}
the analysis of the $z \to 0$ limit unravels the following behavior
\begin{align}
\mathcal{W}^{\rm\bf 6}_{\rm hhh} |_{m_{\rm h} \xi \ll 1}
=
g \mu_{\rm h}^3
\ln \frac{1}{m_{\rm h} \xi}
\left[
\alpha^{\rm\bf 6}_{\rm hhh} \ln \frac{1}{m_{\rm h} \xi} + \beta^{\rm\bf 6}_{\rm hhh} \ln \ln \frac{1}{m_{\rm h} \xi} + \gamma^{\rm\bf 6}_{\rm hhh}
\right]
+ O \left( (m_{\rm h} \xi)^0 \right)
\, ,
\end{align}
where
\begin{align}
\alpha^{\rm\bf 6}_{\rm hhh} \simeq 0.0173 \pm 0.0001
\, , \qquad
\beta^{\rm\bf 6}_{\rm hhh} \simeq - 0.0453 \pm 0.0081
\, , \qquad
\gamma^{\rm\bf 6}_{\rm hhh} \simeq - 0.0141 \pm 0.0202
\, .
\end{align}
We observe that as compared to the singlet case, there is an overall power of the logarithm accompanying the familiar $\xi$-dependence. Thus we anticipate 
the resummation to produce the same functional dependence on $\xi$ up to an extra logarithmic factor which stems from the anomalous dimension\footnote{Notice 
that $X^i$ has a vanishing canonical dimension and therefore does not affect the power-law behavior of the amplitude.} of the two-dimensional bosonic fields $X^i$
($i = 1, \dots, 6$) on the five-sphere which build up the sextet pentagon twist operator $\phi^i_{\pentagon} \sim X^i \phi_{\pentagon}$ which in turn defines the
scalar component of the hexagon Wilson loop in question $\delta^{ij} \mathcal{W}^{\rm 6} \sim \vev{\phi^i_{\pentagon} (\xi) \phi^j_{\pentagon} (0)}$. However, the 
overall normalization will be different and its proper extraction requires resummation. Due to complexity of the asymptotic analysis of multifold integrals and, as
a consequence, the lack of explicit analytical expressions, we performed it numerically. The result of successive additions of multi-scalar exchanges up to five holes 
in shown in Fig.\ \ref{Hseries} (b). The result is fitted by the following formula
\begin{align}
\label{ResummedSextet}
\mathcal{W}^{\rm\bf 6}_\infty
\equiv
\sum_{n = 0}^\infty \, \mathcal{W}^{\rm\bf 6}_{{\rm h} ({\rm hh})^n}  |_{m_{\rm h} \xi \ll 1}
=
g C^{\rm\bf 6}_{\rm h} \, (m_{\rm h} \xi)^{-1/36} \ln^{23/24} \frac{1}{m_{\rm h} \xi}
\, .
\end{align}
with $C^{\rm\bf 6}_{\rm h} \simeq 0.11$. This analysis is in agreement\footnote{We would like to thank Benjamin Basso for bringing the talk \cite{BSVtalk} to our attention 
and useful discussion.} with results announced in Ref.\ \cite{BSVtalk}.

Notice that $\mathcal{W}_{(2,2)}$ in the superloop \re{NMHVhexagon} does not require a dedicated study since it is determined by the sextet multi-scalar 
exchanges we have just discussed,
\begin{align}
\mathcal{W}_{(2,2)} = \sum_{n = 0}^\infty \, \mathcal{W}^{\rm\bf 6}_{{\rm h} ({\rm hh})^n} 
\, .
\end{align}

\subsection{Gluon-scalars}

Finally, we address the component $\mathcal{W}_{(4,0)}$.  Making use of Eq.\ \re{HoleGluonPentagon}, it becomes obvious that as in the previous cases of heavy excitations 
accompanying an infinite tower of scalar exchanges, the contribution in question falls into the product of two terms
\begin{align}
\mathcal{W}_{(4,0)} = \mathcal{W}_{\rm g} \sum_{n = 0}^\infty \, \mathcal{W}^{\rm\bf 1}_{({\rm hh})^n} 
\end{align}
with
\begin{align}
\mathcal{W}_{\rm g} =  \int d\mu_{\rm g} (u) \frac{x^+[u] x^- [u]}{g^2}
\, ,
\end{align}
where the NMHV gluon helicity form factor is given by the product of shifted $x^\pm [u] = x [u \pm \ft{i}{2}]$ Zhukowski variables $x [u] = \ft12 (u + \sqrt{u^2 - (2 g)^2})$
and the infinite sum governed in the ultraviolet regime by the right-hand side of Eq.\ \re{SingletSsm}.

\subsection{Asymptotic form of heavy-particle exchanges}

Let us wrap up our discussion by determining the functional form of the heavy flux-tube exchanges at asymptotic values of $\tau$. Starting with the antifermion integral, we can use 
the saddle point approximation that immediately yields
\begin{align}
\mathcal{W}_{\rm \bar{f}} 
&
= - g^2 \int_{\mathbb{R} + i 0} \frac{d \theta}{\pi \sinh (\theta) \sinh (2 \theta)} {\rm e}^{- \tau \cosh (\theta) + i \sigma \sinh (\theta)}
\nonumber\\
&
\simeq 
g^2
{\rm e}^{- \tau}
\sqrt{\frac{\tau}{2 \pi}} {\rm e}^{- \sigma^2/(2 \tau)}
\left[
1 + \frac{\sigma}{\sqrt{2 \tau}} {\rm e}^{\sigma^2/(2 \tau)} \left( {\rm erf} \left(\frac{\sigma}{\sqrt{2 \tau}} \right) - 1 \right)
+
\frac{5}{6} \frac{1}{\tau} + O \left(\frac{1}{\tau^{3/2}} \right)
\right]
\, .
\end{align}
The leading behavior for the fermion contribution, that arises along with the odd number of accompanying scalars, differs from the above at subleading order in $\tau$ only, namely,
\begin{align}
\mathcal{W}_{\rm f} 
&
= - \int_{\mathbb{R} + i 0} \frac{d \theta}{2 \pi \sinh (\theta) \sinh (2 \theta)} {\rm e}^{- \tau \cosh (\theta) + i \sigma \sinh (\theta)} \frac{1}{(1 + \tanh^2(\theta))}
\nonumber\\
&
\simeq
{\rm e}^{- \tau}
\sqrt{\frac{\tau}{8 \pi}} {\rm e}^{- \sigma^2/(2 \tau)}
\left[
1 + \frac{\sigma}{\sqrt{2 \tau}} {\rm e}^{\sigma^2/(2 \tau)} \left( {\rm erf} \left(\frac{\sigma}{\sqrt{2 \tau}} \right) - 1 \right)
+
\frac{11}{6} \frac{1}{\tau} + O \left(\frac{1}{\tau^{3/2}} \right)
\right]
\, .
\end{align}
Finally, as explained in Ref.\ \cite{Belitsky:2015qla}, to properly take the strong coupling limit of gluons, first one has to pass to the Goldstone sheet $u \to u^{\rm G} + i/2 \to u$ with 
$\Im{\rm m} [u] \geq 1/2$ and then, after rescaling the rapidity $u =2 g \widehat{u}$, send $g \to \infty$,
\begin{align}
\mathcal{W}_{\rm G} 
&
=  \int d\mu_{\rm G} (u) \frac{x^+[u]}{x^- [u]}
\simeq
\int d\mu_{\rm G} (u)
=
- 2g \int \frac{d \theta}{\pi \cosh^2 (2 \theta)} {\rm e}^{- \sqrt{2} \tau \cosh (\theta) + i \sqrt{2} \sigma \sinh (\theta)}
\nonumber\\
&\simeq
2 g {\rm e}^{- \sqrt{2} \tau}
\sqrt{\frac{\sqrt{2} \tau}{\pi}} {\rm e}^{- \sigma^2/(\sqrt{2} \tau)}
\left[
1  - \frac{\sqrt{8}}{\tau} + \frac{4 (\sigma^2 + 4)}{\tau^2} + O (\tau^{-3})
\right]
\, .
\end{align}
Here the first line exhibits the helicity-independence of the gauge transition at strong coupling as the NMHV helicity form factor is $1$ to leading
order in $1/g$ expansion. The same applies to bound states of $\ell$ gauge fields, whose contribution differs from the above consideration by the 
introducing the shifts $\pm i \ell/2$ in Zhukowski variables compared to $\ell = 1$ for a single gluon. The leading order expression is unaffected by these.

\section{Conclusions}

In this work we extended the strong coupling analysis of NMHV hexagon to include hole excitations. The latter develop a nonperturbative regime compared to
all other excitations with their mass gap being exponentially suppressed in strong coupling. Each individual contribution develops logarithmic dependence on
the dimensionless scale $m_{\rm h} z$ which calls for an all-order resummation of all multi-hole exchanges. In all NMHV components the latter factorize into
a multiplier that can be addressed separately from the accompanying heavy flux-tube excitation at leading order in strong coupling. There are two of
these with either singlet or sextet quantum numbers with respect to the internal symmetry group of the parent theory. While the one corresponding to the singlet
was addressed before, presently we added the latter to complete the consideration. The resumed expression was inspired by the reinterpretation of the 
pentagon form factor series in terms of the correlation functions of twist operators in the O(6) sigma model. Like in the MHV case \cite{Basso:2014jfa}, we observed 
a nonperturbative enhancement of the classical area prediction $\exp (- 2 g A_6)$ by multiplicative factors
\begin{align}
\mathcal{W}_\infty^{\bf r} =\left[ 8^{-1/4} \Gamma (\ft54) \right]^{1/36} \xi^{- 1/36} {\rm e}^{\pi g/36}
\left[
C_{\rm h}^{\bf 1} (\pi g)^{- 7/144}\delta_{\bf r, 1}
+
\frac{1}{\pi} C_{\rm h}^{\bf 6} (\pi g)^{281/144}\delta_{\bf r, 6}
\right]
\, , 
\end{align}
depending on the representation of the exchanged scalars.

It is important to find a way to predict the normalization constants in the ultraviolet limit analytically. The fact that the coefficients of the $\xi$-dependence in
individual multi-hole exchanges are given by transcendental numbers suggests that direct resummation is presumably not  the right way to approach this problem
and therefore begs for more efficient techniques. It would be interesting to rephrase these results in a form of Thermodynamic Bethe Ansatz equations similar to the 
ones developed for the MHV amplitudes in Refs.\ \cite{Alday:2009dv,Alday:2010vh}. Apart from that, the overall normalization receives corrections from heavy modes and
inverse coupling expansion. One can extend our current considerations to higher polygons and establish constraints that follow from the Descent Equation 
\cite{CaronHuot:2011kk,Bullimore:2011kg} along the lines of Ref.\ \cite{Belitsky:2015kda} for subleading corrections. 

\section*{Acknowledgments}

We are indebted to Hank Lamm for help with numerical calculations. We would like to thank Benjamin Basso for informing us about the ongoing analysis being 
done in collaboration with Amit Sever and Pedro Vieira \cite{BSVtalk}, which overlaps with the current consideration, as well as discussions at the final stage of this 
work completed during the visit at ENS (Paris) and IPhT (Saclay). We would like to thank Benjamin Basso and Gregory Korchemsky and for the warm hospitality at
respective institutions. This research was supported by the U.S. National Science Foundation under the grants PHY-1068286 and PHY-1403891.

\appendix

\section{Flux-tube equations}
\label{FluxTubeEqsAppendix}

Let us rewrite the flux-tube equations in a form suitable for analysis at strong coupling. Their generic representation for the parity-even and parity-odd cases read
\cite{Basso:2010in,Basso:2013pxa,Belitsky:2014sla}
\begin{align}
\label{EvenEven}
\int_0^\infty
\frac{dt}{t} J_{2n} (2 g t) \left[ \frac{\gamma^{\star}_{u, +} (2 g t)}{1 - {\rm e}^{- t}} - \frac{\gamma^{\star}_{u, -} (2 g t)}{{\rm e}^{t} - 1}  \right]
&
=
\kappa^{\star}_{2 n} (u)
\, , \\
\label{EvenOdd}
\int_0^\infty
\frac{dt}{t} J_{2n - 1} (2 g t) \left[ \frac{\gamma^{\star}_{u, -} (2 g t)}{1 - {\rm e}^{- t}} + \frac{\gamma^{\star}_{u, +} (2 g t)}{{\rm e}^{t} - 1}  \right]
&
=
\kappa^{\star}_{2 n - 1} (u)
\, ,
\end{align}
and
\begin{align}
\label{OddEven}
\int_0^\infty
\frac{dt}{t} J_{2n} (2 g t) \left[ \frac{\widetilde\gamma^{\star}_{u, +} (2 g t)}{1 - {\rm e}^{- t}} + \frac{\widetilde\gamma^{\star}_{u, -} (2 g t)}{{\rm e}^{t} - 1}  \right]
&
=
\widetilde\kappa^{\star}_{2 n} (u)
\, , \\
\label{OddOdd} 
\int_0^\infty
\frac{dt}{t} J_{2n - 1} (2 g t) \left[ \frac{\widetilde\gamma^{\star}_{u, -} (2 g t)}{1 - {\rm e}^{- t}} - \frac{\widetilde\gamma^{\star}_{u, +} (2 g t)}{{\rm e}^{t} - 1}  \right]
&
=
\widetilde\kappa^{\star}_{2 n - 1} (u)
\, ,
\end{align}
respectively. Here the sources depend on the $\star$-type of excitations under consideration. In what follows, we only need the ones corresponding to scalars. However, 
since they will be defined implicitly in our subsequent formulas, we will not display the explicit form in order to save space. In addition, for future reference, we recall the form 
of inhomogeneities for the flux-tube vacuum which read
\begin{align}
\label{VacSource}
\kappa^{\o}_{n} = 2 g \delta_{n,0}
\, , \qquad
\widetilde\kappa^{\o}_{n} = 0
\, .
\end{align}

Following \cite{Basso:2008tx1}, we introduce a functional transformation
\begin{align}
\label{gammaToGamma}
\Gamma^{\rm f}_u (\tau) 
&
\equiv 
\Gamma^{\rm f}_{+,u} (\tau) + i \Gamma^{\rm f}_{-,u} (\tau) 
=
\left(
1 + i \coth \frac{\tau}{4 g}
\right)
\gamma^{\rm f}_{u} (\tau)
\, ,
\\
\label{gammaToGammaTilde}
\widetilde\Gamma^{\rm f}_u (\tau) 
&
\equiv 
\widetilde\Gamma^{\rm f}_{+,u} (\tau) - i \widetilde\Gamma^{\rm f}_{-,u} (\tau) 
=
\left(
1 + i \coth \frac{\tau}{4 g}
\right)
\widetilde\gamma^{\rm f}_{u} (\tau)
\, ,
\end{align}
that has the advantage of removing the explicit dependence on the coupling constant from the Eqs.\ \re{EvenEven} -- \re{OddOdd}. Further, using the Jacobi-Anger summation formula and 
the identity
\begin{align*}
\int_0^\infty \frac{dt}{t} J_0 (2 g t) (\cos (u_1 t) - 1) = 0
\, ,
\end{align*}
($|u_1| < 2 g$) for $\Gamma$, we can cast flux-tube equations for the hole into the form
\begin{align}
\label{GammaH2cos}
\int_0^\infty \frac{dt}{t} ( \cos(u_1 t) - 1 )  
\left[
\Gamma^{\rm h}_{-, u_2} (2 g t)
+
\Gamma^{\rm h}_{+, u_2} (2 g t)
\right]
&
=
-
\int_0^\infty \frac{dt}{t} ( \cos(u_1 t) - 1 )  
\frac{\cos(u_2 t) - {\rm e}^{t/2} J_0 (2 g t)}{\sinh \frac{t}{2}}
\, , \\
\label{GammaH1sin}
\int_0^\infty \frac{dt}{t} \sin(u_1 t)
\left[
\Gamma^{\rm h}_{-, u_2} (2 g t)
-
\Gamma^{\rm h}_{+, u_2} (2 g t)
\right]
&
=
-
\int_0^\infty \frac{dt}{t} \sin(u_1 t) 
\frac{\cos(u_2 t) - {\rm e}^{- t/2} J_0 (2 g t)}{\sinh \frac{t}{2}}
\, ,
\end{align}
and
\begin{align}
\label{GammaHtilde2} 
\int_0^\infty \frac{dt}{t} ( \cos(u_1 t) - 1 )  
\left[
\widetilde\Gamma^{\rm h}_{+, u_2} (2 g t)
-
\widetilde\Gamma^{\rm h}_{-, u_2} (2 g t)
\right]
&
=
-
\int_0^\infty \frac{dt}{t} ( \cos(u_1 t) - 1 )  
\frac{\sin(u_2 t)}{\sinh \frac{t}{2}}
\, , \\
\label{GammaHtilde1}
\int_0^\infty \frac{dt}{t} \sin(u_1 t)
\left[
\widetilde\Gamma^{\rm h}_{+, u_2} (2 g t)
+
\widetilde\Gamma^{\rm h}_{-, u_2} (2 g t)
\right]
&
=
-
\int_0^\infty \frac{dt}{t} \sin(u_1 t) 
\frac{\sin(u_2 t)}{\sinh \frac{t}{2}}
\, .
\end{align}
These results are used in the main text.

\section{Exchange relations}

To partially verify our findings for nonperturbative corrections derived in the main text, we will rewrite the energy and momentum of the hole, which are
are conventionally expressed via the vacuum flux-tube function $\gamma^{\o}(t)$ \cite{Basso:2010in}
\begin{align}
E_{\rm h} (u)
&=
1 + \int_0^\infty \frac{dt}{t} \frac{\gamma^{\o} (- 2 g t)}{{\rm e}^t - 1} \left( {\rm e}^{t/2} \cos (u t) - J_0 (2 g t) \right)
\, , \\
p_{\rm h} (u)
&=
2 u - \int_0^\infty \frac{dt}{t} \frac{\gamma^{\o} (2 g t)}{{\rm e}^t - 1} {\rm e}^{t/2} \sin (u t) 
\, ,
\end{align}
in terms of the hole flux-tube functions $\gamma^{\rm h}_u$ and $\widetilde\gamma^{\rm h}_u$. Let us demonstrate it for the momentum and just quote the final answer for the energy.

To start with, let us recall that the even and odd components of the flux-tube functions are entire functions and admit convergent Neumann expansions in terms 
of Bessel functions. Then one can write the above formula in the form of an infinite series representation
\begin{align}
p_{\rm h} (u) 
=
2 u
+
2 \sum_{n \geq 1} (2n) \gamma^{\o}_{2n} \widetilde\kappa^{\rm h}_{2n} (u)
+
2 \sum_{n \geq 1} (2n - 1) \gamma^{\o}_{2n - 1} \widetilde\kappa^{\rm h}_{2n - 1} (u)
\, ,
\end{align}
making use of the sources defining inhomogeneities in the flux-tube equations for scalars. Multiplying Eqs.\ \re{OddEven} and \re{OddOdd} by $2 (2n) J_{2n} (2 g t)$ 
and $2 (2n - 1) J_{2n - 1} (2 g t)$, respectively, and  summing over positive values of $n$, we find for their sum
\begin{align}
2 \sum_{n \geq 1} (2n) \gamma^{\o}_{2n} \widetilde\kappa^{\rm h}_{2n} (u)
+
2 \sum_{n \geq 1} (2n - 1) \gamma^{\o}_{2n - 1} \widetilde\kappa^{\rm h}_{2n - 1} (u)
&
=
\int_0^\infty \frac{dt}{t} 
\bigg[
\frac{\gamma^{\o}_+ (2 g t) \widetilde\gamma^{\rm h}_{+, u} (2 g t) + \gamma^{\o}_- (2 g t) \widetilde\gamma^{\rm h}_{-, u} (2 g t) }{1 - {\rm e}^{- t}}
\nonumber\\
&\qquad\quad
+
\frac{\gamma^{\o}_+ (2 g t) \widetilde\gamma^{\rm h}_{-, u} (2 g t) - \gamma^{\o}_- (2 g t) \widetilde\gamma^{\rm h}_{+, u} (2 g t) }{{\rm e}^{t} - 1}
\bigg]
. 
\end{align}
Now, expanding the hole flux-tube functions in the Neumann series provides gives a very concise representation of the right-hand side
\begin{align}
2 \sum_{n \geq 1} (2n) \widetilde\gamma^{\rm h}_{2n} (u) \kappa^{\o}_{2n}
+
2 \sum_{n \geq 1} (2n - 1) \widetilde\gamma^{\rm h}_{2n - 1} (u) \kappa^{\o}_{2n - 1}
=
4 g \widetilde\gamma^{\rm h}_1 (u)
\, ,
\end{align}
where we employed the explicit form of the sources for the vacuum \re{VacSource}.

Analogous consideration can be done for the energy such that one can rewrite the dispersion relation in the form
\begin{align}
\label{HoleEandPfromGammaH}
E_{\rm h} (u) 
= 
1 + 2 \lim_{t \to 0} \frac{\gamma_u^{\rm h} (2 g t)}{t}
\, , \qquad
p_{\rm h} (u) 
=
2 u + 2 \lim_{t \to 0} \frac{\widetilde\gamma_u^{\rm h} (2 g t)}{t}
\, .
\end{align}
Here we relied on the fact that only the leading term in the Neumann expansion induces a nontrivial contribution (with subleading ones scaling as powers of $t$ which
vanish in the limit in question). These agree with Ref.\ \cite{Basso:2013pxa}.


\end{document}